\documentclass[11pt]{article}

\newcommand{\Slash}[1]{\ooalign{\hfil/\hfil\crcr$#1$}}
\newcommand{\be}{\begin{equation}}
\newcommand{\ee}{\end{equation}}

\usepackage[english]{babel}
\usepackage{amsmath}
\usepackage{amssymb}
\usepackage{amsfonts}
\usepackage{amscd}
\usepackage{color}
\usepackage{graphicx}
\usepackage[pdftex]{hyperref}
  \newcommand{\cH}{{\cal H}}

  \newcommand{\cP}{{\cal P}}

\newcommand{\bP}{{\mathbf P}}  
\newcommand{\bR}{{\mathbf R}}

\newcommand{\Nc}{N_{\rm c}}
\newcommand{\Nf}{N_{\rm f}}

\newcommand{\bea}{\begin{eqnarray}} \newcommand{\eea}{\end{eqnarray}}
\newcommand{\beann}{\begin{eqnarray*}}  \newcommand{\eeann}{\end{eqnarray*}}
\newcommand{\bfig}{\begin{figure}} \newcommand{\efig}{\end{figure}}
\newcommand{\ba}{\begin{array}} \newcommand{\ea}{\end{array}}
\newcommand{\bcen}{\begin{center}} \newcommand{\ecen}{\end{center}}
\newcommand{\btab}{\begin{tabular}} \newcommand{\etab}{\end{tabular}}

\def\tr{\operatorname{tr}}     \def\Tr{\operatorname{Tr}}

\allowdisplaybreaks[4]

\date{}
\begin{document}

\begin{flushright}
\today\\
TAUP-2941/11
\end{flushright}

\vspace{0.1cm}

\begin{center}
  {\LARGE

Holographic realization of large-$\Nc$ orbifold equivalence
\\with non-zero chemical potential

  }
\end{center}
\vspace{0.1cm}
\vspace{0.1cm}
 \begin{center}
    Masanori Hanada${}^{a,b}$,\footnote {E-mail address: hanada@post.kek.jp}
    Carlos Hoyos${}^{a,c}$,\footnote{E-mail address: choyos@post.tau.ac.il}
    Andreas Karch${}^a$\footnote{E-mail address: akarch@uw.edu}
    and Laurence G. Yaffe${}^a$\footnote{E-mail address: lgy@uw.edu}
\\[0.6cm]
${}^a${\it Department of Physics, University of Washington, Seattle, WA 98915-1560, USA}\\
${}^b${\it KEK Theory Center, High Energy Accelerator Research Organization (KEK), \\
Tsukuba 305-0801, Japan} \\
${}^c${\it Raymond and Beverly Sackler School of Physics and Astronomy, Tel-Aviv University, \\
Ramat-Aviv 69978, Israel}
\end{center}
\vspace{1.5cm}

\begin{center}
  {\bf Abstract}
\end{center}

Recently, it has been suggested that large-$\Nc$ orbifold equivalences may
be applicable to certain theories with chemical potentials, including QCD,
in certain portions of their phase diagram.
When valid, such an equivalence offers the possibility of relating
large-$\Nc$ QCD at non-zero baryon chemical potential,
a theory with a complex fermion determinant,
to a related theory whose fermion determinant is real and positive.
In this paper, we provide a test of this large $\Nc$ equivalence using
a holographic realization of a supersymmetric theory with baryon chemical
potential and a related theory with isospin chemical potential.
We show that the two strongly-coupled, large-$\Nc$ theories
are equivalent in a large region of the phase diagram.

\newpage

\tableofcontents

\section{Introduction}
\hspace{0.51cm}

Understanding QCD at non-zero baryon density is an important goal,
both for its intrinsic interest and for applications such as the
structure of neutron stars and the mechanism of core-collapse supernova.
Due to the notorious \textit{sign problem},\footnote
    {%
    More properly, this should be called a \textit{phase} problem.
    See Appendix~\ref{app:sign} for a brief summary.
    }
we lack generally effective methods for performing
numerical simulations of gauge theories
with a baryon chemical potential.
When a non-zero baryon number chemical potential is present, the determinant
of the Euclidean Dirac operator is no longer positive
and standard Markov-chain Monte-Carlo methods are not applicable.
Although many schemes have been proposed to address the sign problem
\cite{Parisi:1984cs,Klauder:1983sp,Aarts:2009uq,Fodor:2001pe,Anagnostopoulos:2001yb,
Anagnostopoulos:2010ux,Allton:2002zi},
it is fair to say that no fully satisfactory solution has been found.
At the same time, condensed matter phases of several QCD-like theories
which do not suffer from the sign problem have been studied numerically,
in the hope that one may extract lessons about strongly interacting
finite density systems which will also apply to QCD at finite baryon density.
Examples include $SU(2)$ Yang-Mills (YM) with even numbers of fundamental
flavors \cite{Kogut:1999iv,Kogut:2000ek},
$SU(\Nc)$ YM with adjoint fermions \cite{Kogut:2000ek},
and QCD with an isospin chemical potential \cite{Alford:1998sd,Son:2000xc}.
However, there is no solid argument delineating the extent to which these
theories can reproduce properties of QCD with a baryon chemical potential.

In recent years, it has been understood that a network of large-$\Nc$
equivalences relate various non-Abelian gauge theories with differing gauge
groups and matter content
\cite{Bershadsky:1998cb,Bershadsky:1998mb,Kachru:1998ys,
Kovtun:2003hr,Armoni:2003gp}.
These equivalences, which are generated by appropriate orbifold projections,
relate the leading large $\Nc$ behavior of connected correlators
of specific classes of observables.
The large-$\Nc$ equivalences are valid provided certain symmetry
realizations are satisfied \cite{Kovtun:2004bz}.
For example, $SU(\Nc)$ and $SO(2\Nc)$ Yang-Mills theories have
coinciding large $\Nc$ limits of all Wilson loop expectation values
(as well as connected correlators of real parts of Wilson loops),
provided charge conjugation symmetry is not spontaneously broken
in the $SU(\Nc)$ theory \cite{Unsal:2006pj}.

\begin{figure}
\small
\begin{center}
\begin{picture}(280,150)
\put(000,20){\shortstack
    {$U(\Nc)$\\$\Nf$ fundamentals\\[2pt]baryon chemical potential $\mu_B$}}
\put(160,20){\shortstack
    {$U(\Nc)$\\$\Nf$ fundamentals\\[2pt]isospin chemical potential $\mu_I$}}
\put(105,120){\shortstack
    {$O(2\Nc)$\\$\Nf$ fundamentals\\[2pt]chemical potential $\mu_F$}}
\put(110,105){\vector(-1,-1){40}}
\put(185,105){\vector( 1,-1){40}}
\end{picture}
\end{center}
\caption{%
    An orbifold projection acting on a parent
    $O(2\Nc)$ Yang-Mills theory with $\Nf$ fundamental representation
    Dirac fermions and a flavor-singlet chemical potential $\mu_F$
    may generate a $U(\Nc)$ daughter theory with $\Nf$
    fermions and a baryon chemical potential $\mu_B$ (left)
    or, provided $\Nf$ is even, the same $U(\Nc)$ theory with
    an isospin chemical potential $\mu_I$ (right)
    \cite{Cherman:2010jj}.
    In the parent $O(2\Nc)$ theory with even $\Nf$,
    there is no distinction between
    a baryon or isospin chemical potential.
    }\label{fig:Projection_nonSUSY}
\end{figure}
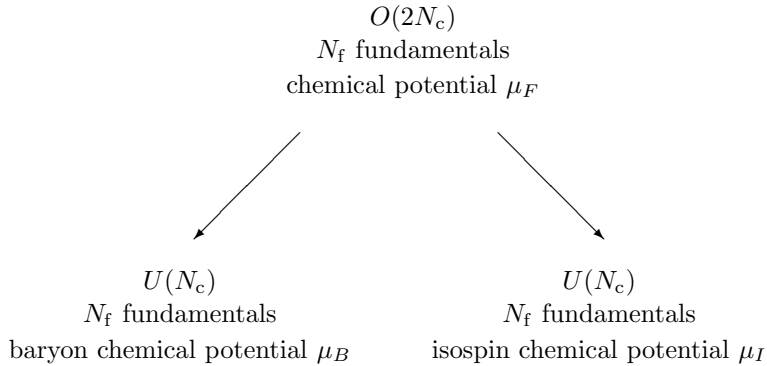

We will be concerned with QCD-like theories containing
fundamental representation fermions and non-zero chemical potentials.
Specifically, we will discuss:
\begin{enumerate}
\item
    $SO(2\Nc)$ Yang-Mills with $\Nf$ Dirac fundamental representation
    fermions and a non-zero fermion chemical potential
    $\mu_F$, under which all $\Nf$ flavors have charge $+1$.
    For brevity, we will denote this theory as $SO(2\Nc)_{F}$.
\item
    $SU(\Nc)$ Yang-Mills with $\Nf$ fundamental representation fermions
    and a non-zero baryon chemical potential $\mu_B$, under which
    all $\Nf$ fermion flavors have charge $+1$.
    For brevity, we will denote this theory as QCD$_{B}$.
\item
    $SU(\Nc)$ Yang-Mills with $\Nf$ fundamental representation fermions,
    with $\Nf$ even,
    and a non-zero isospin chemical potential $\mu_I$, under which
    half the fermion flavors have charge $+1$ and half have charge $-1$.
    For brevity, we will denote this theory as QCD$_{I}$.
\end{enumerate}
Although QCD$_B$ suffers from a sign problem, this is not the case
for either QCD$_I$ or $SO(2\Nc)_F$, as both of these theories have
a real and positive fermion determinant \cite{Alford:1998sd,Cherman:2010jj}.

As figure~\ref{fig:Projection_nonSUSY} schematically depicts,
starting from the $SO(2\Nc)_F$ theory one choice of
orbifold projection yields QCD$_B$,
while a different choice yields QCD$_I$.%
\footnote{
    Strictly speaking, the orbifold projection maps a parent theory
    with $SO(2\Nc)$ gauge group to a daughter $U(\Nc)$ gauge theory.
    But the difference between $U(\Nc)$ and $SU(\Nc)$ theories is
    sub-dominant in the large $\Nc$ limit.
    Note that in a $U(\Nc)$ theory with chemical potential, the
    $U(1)$ part of the gauge field is to be fixed at infinity.
    In finite volume, the theory should be defined with Dirichlet boundary
    conditions on the $U(1)$ gauge field, not periodic.
    (In practice, for lattice simulations, it is more convenient to simply
    use the $SU(\Nc)$ theory.)
    }
Based on this observation,
it has recently been suggested that large $\Nc$ equivalences
may relate suitable observables in the parent
$SO(2\Nc)_F$ theory to corresponding observables
in either QCD$_B$ or QCD$_I$
\cite{Cherman:2010jj,Hanada:2011ju,Cherman:2011mh}.
In portions of the phase diagram where both equivalences are valid
(if such regions exist),
this implies that one may obtain quantitative information
about large-$\Nc$ QCD with a baryon chemical potential from studies
of the same theory with an isospin chemical potential,
thereby circumventing the sign problem.%
\footnote
    {%
    Readers should refer to section 2, and
    refs.~\cite{Cherman:2010jj,Hanada:2011ju,Cherman:2011mh},
    for more details and more nuanced discussion.
    }

When $\Nc \to \infty$ with $\Nf$ fixed,
a comparison of planar Feynman diagrams
in the parent $SO(2\Nc)$ and daughter $SU(\Nc)$ theories
shows that they coincide
\cite{Bershadsky:1998cb,Bershadsky:1998mb,Cherman:2010jj,Hanada:2011ju}. 
(For other approaches see \cite{Kovtun:2003hr}.) 
This analysis is unaffected by the presence of a non-zero chemical potential.
Coinciding perturbative expansions does not, however, necessarily imply
a genuine non-perturbative equivalence.
Necessary conditions for a valid equivalence include a requirement that
the symmetries used to define the orbifold projection not be
spontaneously broken in the parent theory \cite{Kovtun:2003hr}.
Since the projection leading to QCD$_B$
is generated by a combination of a gauge transformation and a $U(1)_F$
phase rotation, this projection can only lead to a valid large-$\Nc$
equivalence in portions of the phase diagram where the $U(1)_F$
global symmetry associated with net fermion number is unbroken.
In other words, a large $\Nc$ equivalence relating QCD with
baryon and isospin chemical potentials can only apply to portions
of the phase diagram in which fermions do not condense to form a superfluid.
In simpler examples, analogous conditions on symmetry realizations are
both necessary and sufficient conditions for the validity of large $\Nc$
equivalences \cite{Kovtun:2003hr};
whether this is the case in the present example is not yet clear.

In this paper, we use gauge/gravity duality to test the validity of analogous
possible large $\Nc$ equivalences relating supersymmetric generalizations
of the above theories.
Although the trio of theories we consider will not include QCD itself,
the arguments of 
refs.~\cite{Cherman:2010jj,Hanada:2011ju,Cherman:2011mh}
are equally applicable to the supersymmetric theories we consider.
By considering supersymmetric theories, and using holographic methods,
it will be possible to examine relations between theories with different
types of chemical potential directly in the limit of strong coupling
(and large $\Nc$) using simple analytic methods.
We will find that, in a large region of the phase diagram with
no spontaneous breaking of flavor symmetries, large $\Nc$ equivalences
between our theories are valid.

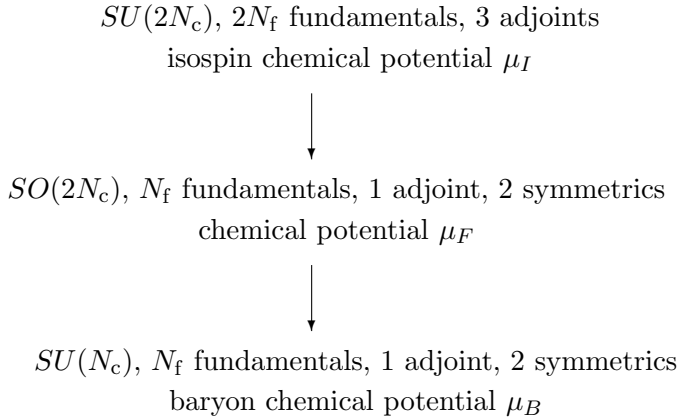
\begin{figure}
\begin{center}
\begin{picture}(200,160)
\put(10,140){\shortstack
    {$SU(2\Nc)$, $2\Nf$ fundamentals, 3 adjoints\\[2pt]isospin chemical potential $\mu_I$}}
\put(-25,75){\shortstack
    {$SO(2\Nc)$, $\Nf$ fundamentals, 1 adjoint, 2 symmetrics\\[2pt]chemical potential $\mu_F$}}
\put(-15,10){\shortstack
    {$SU(\Nc)$, $\Nf$ fundamentals, 1 adjoint, 2 symmetrics\\[2pt]baryon chemical potential $\mu_B$}}
\put(90,130){\vector(0,-1){25}}
\put(90,65){\vector(0,-1){25}}
\end{picture}
\end{center}
\caption{
    ${\cal N}=1$ supersymmetric theories related by orbifold projections.
    In the $SO(2\Nc)$ theory, there is no distinction between
    a baryon or isospin chemical potential.
    }\label{fig:Projection_SYM}
\end{figure}

Our holographic construction involves an orbifold and an orientifold
projection of the D3/D7-system, with $\Nc$ D3 branes, $\Nf$ D7
branes, and always $\Nf\ll \Nc$.
At large-$\Nc$
and strong 't Hooft coupling, the low energy theory on
the D3 branes is described by classical type IIB supergravity in
$AdS_5\times S^5$, with probe D7 branes wrapping an $S^3$ in the
$S^5$ \cite{Karch:2002sh}.
The projections act on the geometry,
changing it to $AdS_5\times {\bf RP}^5$. An isospin chemical potential
in the original theory is described by a particular configuration
of the gauge field on the D7 brane. We show that after the projection
the isospin chemical potential becomes a baryon chemical potential.
Figure~\ref{fig:Projection_SYM} illustrates the connections between
the corresponding field theories.
We prove that, provided the projection symmetries are not broken,
the equations of motion of the D7 branes coincide in both theories.
As we discuss,
this implies that the conjectured large $\Nc$ equivalences are valid
in these theories in those regions of the phase diagram where flavor
symmetries are unbroken and no additional fields become active.

The paper is organized as follows: in \S\,\ref{sec:QCD} we review
in more detail the proposed equivalences involving large-$\Nc$ QCD.
In \S\,\ref{sec:holography} we provide a holographic realization
for our supersymmetric generalization.
Finally, in \S\,\ref{sec:conclusions} we discuss the regime of
validity of the equivalence, and the consequences for the phase diagram.

\section{Orbifold projections for large-$\Nc$ QCD}\label{sec:QCD}
\hspace{0.51cm}
Consider an $SO(2\Nc)$ Yang-Mills theory coupled to $\Nf$ fundamental
representation Dirac fermions.
The Lagrange density is
\begin{eqnarray}
\label{eq:SOLagrangian}
\mathcal{L_{\rm SO}}
=\frac{1}{4 g_{SO}^{2} }\, \tr F_{\mu \nu}^2
+ \sum_{a =1}^{\Nf} \>
\bar{\psi}_{a} \left(\gamma^{\mu} D_{\mu} + m_{q}+\mu_{F}\,\gamma^{0}\right)
\psi_{a}
\,,
\end{eqnarray}
where $F_{\mu \nu}$ is the field strength of the $SO(2\Nc)$
gauge field $A_{\mu}$, $D_{\mu} \equiv \partial_{\mu}+ A_{\mu}$,
$\psi_{a}$ is a Dirac fermion in the vector representation of $SO(2\Nc)$,
and $m_{q}$ and $\mu_{F}$ are the quark mass and fermion chemical potential,
respectively.
Because the gauge field is  real, the Dirac operator
$\mathcal D \equiv (\gamma^{\mu} D_{\mu} + m_{q}+\mu_{F}\,\gamma^{0})$ satisfies
$(C\gamma_5)D\, (C\gamma_5)^{-1}=D^\ast$, where $C$ is the charge
conjugation matrix defined by $C\gamma_\mu C^{-1}=-\gamma_\mu^T=-\gamma_\mu^\ast$.%
\footnote
{%
We use $({+}{+}{+}{+})$ metric signature and Hermitian gamma matrices.
}
If $v$ is an eigenvector of the Dirac operator $\mathcal D$
with an eigenvalue $\lambda$, 
$\mathcal D\,v=\lambda v$,
then $(C\gamma_5)^{-1}v^\ast$ is another eigenvector of $\mathcal D$
with eigenvalue $\lambda^\ast$, and is linearly independent of $v$
even when $\lambda$ is real. (See \S\,2.3 of ref.~\cite{Hands:2000ei}.)
Therefore the determinant of $D$ is always real and positive, implying that
standard Markov chain Monte-Carlo simulation techniques
may be used \cite{Cherman:2010jj}.

When $m_{q} = \mu_{F}=0$, the Lagrangian \eqref{eq:SOLagrangian} has a
manifest $SU(\Nf)_{L}\times SU(\Nf)_{R} \times U(1)_{F} \times U(1)_{A}$
flavor symmetry,
just like $SU(\Nc)$ QCD.
However, the flavor symmetry of the theory is actually larger than this
due to the fact that $SO(2\Nc)$ is a real gauge group;
classically it extends to $U(2\Nf)$~\cite{Coleman:1980mx,Peskin:1980gc}.
The axial $U(1)_A \subset U(2\Nf)$ is anomalous, and at the
quantum level the (continuous part of the) flavor symmetry is
$SU(2\Nf)$, which spontaneously breaks to
$SO(2\Nf) \supseteq SU(\Nf)_{V} $ due to
the formation of a chiral condensate $\langle\bar{\psi}{\psi}\rangle$.
The resulting massless Nambu-Goldstone bosons
span the $SU(2\Nf)/SO(2\Nf)$ coset space.
In contrast to QCD, some of these Nambu-Goldstone bosons, which we
will refer to as baryonic pions, are charged under $U(1)_{F}$.
Ordinary pions are created by operators that look like
$\bar{\psi}_a \gamma_{5} \psi_b$,
while baryonic pions are created by color-singlet operators of the form
$\psi_{a}^{T} C\gamma_5  \psi_{b}$ and $\bar{\psi}_{a} C\gamma_5  \bar{\psi}^T_{b}$.

\subsection{From $SO(2\Nc)_F$ to QCD with a baryon chemical potential}
\hspace{0.51cm}
To perform an orbifold projection, one identifies a discrete subgroup
of the symmetry group of the parent theory, which for us is the
$SO(2\Nc)_F$ theory, and then removes all of the degrees of freedom
in the parent theory which are not invariant under the chosen discrete
symmetry.
This yields a daughter theory, which will turn out
to be large-$\Nc$ QCD.

The required orbifold projection is a
$\mathbb{Z}_{2}$ subgroup of the $SO(2\Nc) \mbox{ gauge} \times U(1)_{F}$ flavor
symmetry of the $SO(2\Nc)_F$ theory.
To define the orbifold projection, which we will denote as $\cP_B$,
take $J_{2\Nc}\in SO(2\Nc)$ to be given by
$J_{2\Nc} = i\sigma_{2} \otimes \textbf{1}_{\Nc}$,
where $\textbf{1}_{N}$ denotes an $N \times N$ identity matrix.
The group element $J_{2\Nc}$ generates a $\mathbb{Z}_{4}$ subgroup
of $SO(2\Nc)$.
Next, let $\omega = e^{i \pi/2} \in U(1)_{F}$ denote the phase
which generates a $\mathbb{Z}_{4}$ subgroup of $U(1)_{F}$.
The discrete symmetry which will define the orbifold projection acts on
on the fields $A_{\mu}$, $\psi_{a}$ as
\begin{equation}
A_{\mu} \to J_{2 \Nc} A_{\mu} J_{2 \Nc}^{-1}, \qquad
\psi_{a} \to \omega J_{2 \Nc} \psi_{a} \,.
\end{equation}
Since $J_{2 \Nc}^2=-\textbf{1}_{2 \Nc}$,
and $(\omega J_{2\Nc})^2 = +\textbf{1}_{2 N_c}$,
this symmetry transformation
generates a $\mathbb{Z}_{2}$ subgroup of $SO(2\Nc)\times U(1)_{F}$.


The action of the orbifold projection on the basic fields is
\begin{equation}\label{eq:qcdbarproj}
\cP_B A_\mu=\tfrac{1}{2}\left(A_\mu + J_{2 \Nc} A_{\mu} J_{2 \Nc}^{-1}\right),
\qquad
\cP_B \psi = \tfrac{1}{2}\left(\psi+J_{2 \Nc} \psi K_{\Nf}^{-1}\right)
\end{equation}
where, for later convenience,
we have defined a matrix $K_{\Nf}^{-1}\equiv i\textbf{1}_{\Nf}$
acting on flavor indices.
To display the action of the projection more explicitly,
it is convenient to block-decompose the gauge and fermion fields.
The gauge field $A_{\mu}$ may be written in terms of four
$\Nc\times \Nc$ blocks as
\begin{align}
A_\mu
\equiv
\left(
\begin{array}{cc}
A_\mu^A+B_\mu^A & C_\mu^A-D_\mu^S\\
C_\mu^A+D_\mu^S & A_\mu^A-B_\mu^A
\end{array}
\right),
\end{align}
where fields marked with an `$A$' or `$S$' superscript are anti-symmetric
or symmetric matrices, respectively.
Under the $\mathbb{Z}_{2}$ symmetry transformation (\ref{eq:qcdbarproj}),
$A_\mu^A$, and $D_\mu^S$ are even while $B_\mu^A$, and $C_\mu^A$ are odd,
so the orbifold projection sets $B_{\mu}^{A} = C_{\mu}^{A} = 0$.  Hence
\begin{align}
\cP_B A_{\mu}
=
\left(
\begin{array}{cc}
A_{\mu}^A  & -D_\mu^S\\
D_\mu^S & A_\mu^A
\end{array}
\right).
\end{align}
If one defines the unitary matrix
\begin{eqnarray}\label{eq:P}
P = \frac{1}{\sqrt{2}}\left(
\begin{array}{cc}
\textbf{1}_{\Nc} & i \textbf{1}_{\Nc} \\
\textbf{1}_{\Nc} & -i \textbf{1}_{\Nc}
\end{array}
\right),
\end{eqnarray}
then
\begin{eqnarray}
P \, \cP_B A_{\mu} P^{-1} =
  \left(
\begin{array}{cc}
\mathcal{A}_{\mu} & 0\\
0 & -\mathcal{A}_{\mu}^T
\end{array}
\right),
\label{gauge_field_diagonal_basis}
\end{eqnarray}
where $\mathcal{A}_{\mu} \equiv A_{\mu}^{A}+ iD_{\mu}^{S} $ 
is a $U(\Nc)$ gauge field.  At large $\Nc$, we can neglect the difference between $U(\Nc)$ and $SU(\Nc)$ up to $1/\Nc^{2}$ corrections.

We can split the $2\Nc$-component fundamental fermions of the $SO(2 \Nc)$
theory into two $\Nc$-component fields,
\begin{eqnarray}
\psi
=
\left(
\begin{array}{c}
\psi_1\\
\psi_2
\end{array}
\right),
\end{eqnarray}
and then we use the matrix \eqref{eq:P} to change basis.
This yields
\begin{eqnarray}
P\psi
=
\left(
\begin{array}{c}
\psi_+\\
\psi_-
\end{array}
\right),
\end{eqnarray}
where $\psi_\pm\equiv (\psi_1\pm i\psi_2)/\sqrt{2}$.
From eq.~\eqref{gauge_field_diagonal_basis}, one sees that
$\psi_+$ and $\psi_-$ transform as fundamental and antifundamental
representations under $SU(\Nc)$, respectively.
After the projection, only $\psi_+$ survives.

If we take the Lagrangian of the parent theory and apply the orbifold projection, it becomes
\begin{eqnarray}
\mathcal{L} = \frac{1}{4 g_{SU}^{2} } \Tr \mathcal{F}_{\mu \nu}^2
+
\sum_{a=1}^{\Nf} \>
\bar{\lambda}_{a}\left( \gamma^{\mu} {\cal D}_{\mu} + m_q + \mu_B\gamma^4\right)\lambda^{a},
\end{eqnarray}
where $\mathcal{F}_{\mu\nu}$ is the field strength of the $SU(\Nc)$ gauge field
$\mathcal{A}_{\mu} = A^{A}_{\mu} +i D^{S}_{\mu} $,
${\cal D}_{\mu} = \partial_{\mu} + \mathcal{A}_{\mu}$,
$\lambda^{a} = \sqrt{2}\,\psi_{+}^{a}$,
and the gauge coupling is given by $g_{SU}^{2} = g_{SO}^{2}$. 

In the large-$\Nc$ limit for fixed $\Nf$, connected correlation functions of operators
${\cal O}_i^{(p)}$
in the parent $SO$ theory
which are
invariant under the projection symmetry,
and their counterparts ${\cal O}_i^{(d)}$ in the daughter $SU$ theory
which are formed from the projected fields,
coincide to all orders in perturbation theory \cite{Bershadsky:1998cb},
\begin{eqnarray}
\langle{\cal O}_1^{(p)}{\cal O}_2^{(p)}\cdots\rangle_{p}
=
\langle{\cal O}_1^{(d)}{\cal O}_2^{(d)}\cdots\rangle_{d}.
\end{eqnarray}
The baryonic pion fields do not survive the projection,
so there is no equivalent to them in the daughter theory.

\subsection{From $SO(2\Nc)_F$ to QCD with an isospin chemical potential}
\hspace{0.51cm}
When the number of flavors in the parent $SO(2 \Nc)$ theory is even, $\Nf=2k$, it is also possible to define a projection which yields large-$\Nc$ QCD with
an isospin chemical potential.
The projection for the gauge field is the same as in eq.~\eqref{eq:qcdbarproj},
but we now choose a different orbifold action on the flavor indices of the fermions.
Let us write the fermions using $\Nc\times \Nf$-component fields as
\begin{equation}
\psi=\left(\begin{array}{cc} \psi_+^{(1)} & \psi_+^{(2)} \\ \psi_-^{(1)} & \psi_-^{(2)} \end{array} \right)\,.
\end{equation}
In this basis, the orbifold action is
\begin{eqnarray}
\psi\to J_{2 \Nc}\, \psi \, J_{2k}^{-1}.
\label{eq:projection_isospin}
\end{eqnarray}
This transformation also generates a $\mathbb{Z}_2$ group.
The action of the orbifold projection $\cP_I$ is
\begin{equation}\label{eq:isoproj}
\cP_I A_\mu=\tfrac{1}{2}\left(A_\mu + J_{2 \Nc} A_{\mu} J_{2 \Nc}^{-1}\right), \ \ \cP_I \psi = \tfrac{1}{2}\left(\psi+J_{2 \Nc} \psi J_{2 k}^{-1}\right).
\end{equation}
Defining $\varphi_{\pm}=(\psi_{\pm}^{(1)} \mp i \psi_{\pm}^{(2)})/\sqrt{2}$
and $\xi_{\pm}=(\psi_{\pm}^{(1)} \pm \ i \psi_{\pm}^{(2)})/\sqrt{2}$, 
one sees that $\varphi_{\pm}$ survive while $\xi_{\pm}$
is eliminated by the projection (\ref{eq:isoproj}).
Since $\varphi_{+}$ and $\varphi_{-}$ couple to ${\cal A}_{\mu}$ and ${\cal A}_{\mu}^C$, respectively,
the fermionic part of the action of the daughter theory can be written as
\begin{eqnarray}
\sum_{f=1}^{k} \sum_{\pm} \>
\bar{\lambda}_{ \pm}^{(f)}\left( \gamma^{\mu} {D}_{\mu} + m \pm \mu \gamma^4 \right)\lambda_{ \pm}^{(f)},
\end{eqnarray}
where $\lambda_{+}^{(f)}=\sqrt{2}\,\varphi_+^{(f)}$,
$\lambda_{-}^{(f)}=\sqrt{2}\,(\varphi_-^{(f)})^C$,
and we have now written the flavor index $(f) = 1,\cdots,k$ explicitly.
This theory has an isospin chemical potential $\mu_I\equiv 2\mu$.

\subsection{Validity of large-$\Nc$ equivalences and their application to the  sign problem}\label{sec:qcdcomments}
\hspace{0.51cm}
The perturbative proof of the parent-daughter equivalence with isospin chemical potential is valid also when quark loops are included in planar diagrams, so it is possible to extend the analysis to include $\Nf/\Nc$ corrections. However, this is not possible for the projection to a theory with baryon chemical potential. The difference stems from the properties of the projection in the flavor sector, while the projection to isospin chemical potential is performed using a {\em regular} representation \cite{Bershadsky:1998cb,Bershadsky:1998mb}
\begin{equation}\label{eq:regularity}
\tr J_{2 k}=0, \ \ J_{2 k}^2=\pm\textbf{1}_{2 k},
\end{equation}
these conditions are not satisfied for the representation used to do the projection to the theory with a baryon chemical potential, where we have used $K_{2 k}$ instead of $J_{2 k}$. Only diagrams containing a single quark loop produce the same result in parent and daughter theories.

To go beyond the perturbative proof of the equivalence one needs to do a careful analysis of the necessary and sufficient conditions that must be obeyed for the equivalence to hold.
A necessary condition is that the projection symmetry not be spontaneously broken in the parent \cite{Kovtun:2003hr}.
The $U(1)_B$ symmetry, which is used for the projection from the $SO(2 \Nc)$ theory to QCD with a baryon chemical potential,
breaks to ${\mathbb Z}_2$ when the baryonic pion condenses
(\emph{e.g.}, when $\mu>m_\pi/2$ at zero temperature).
Therefore, the parent-daughter equivalence can hold only at smaller values of the chemical potential.%
\footnote
    {%
    Note that the chemical potential at which baryonic pions condense is temperature
    dependent, and should increase with increasing temperature.
    }
On the other hand, the projection symmetry to obtain QCD with isospin chemical potential should not be spontaneously broken for any $\mu$;
in this case condensation of baryonic pions in the parent theory is mapped to
pion condensation in the daughter theory.

Clearly, if it were possible to show that these equivalences hold nonperturbatively, they would be very useful because one would be able to derive properties of a large-$\Nc$ QCD theory with baryonic chemical potential from a $SO(2\Nc)$ theory or from  large-$\Nc$ QCD with isospin chemical potential, both of which are free of the sign problem. This could also explain why the phase quenching approximation in QCD is quite good
 --- for a certain class of operators (\emph{e.g.}, the chiral condensate), the phase quenching approximation becomes exact in the  large-$\Nc$ limit.\footnote{%
Note that (for $\Nf$ even) dropping the phase of the fermion determinant turns the functional integral for QCD$_B$ into that for QCD$_I$.
}
The phase quenching approximation for the chiral condensate is exact
in the chiral random matrix model \cite{Klein:2003fy,Hanada:2011ju}.
The orbifold equivalence, if true, would ensure that the phase quenching
approximation in QCD is exact for a large class of observables in the
large-$\Nc$ limit, even
beyond the parameter region where the chiral random matrix model is valid
(the ``$\epsilon$-regime").

To provide a nonperturbative proof of the orbifold equivalence in QCD with chemical potentials is beyond the scope of this paper. However, in the following section we will show that analogous equivalences hold in a class of supersymmetric cousins of QCD which have gravity duals.

\section{A holographic realization}\label{sec:holography}
\hspace{0.51cm}
It is possible to build a simple supersymmetric model where an isospin chemical potential is projected into a baryon chemical potential. The model is one of the examples mentioned in ref.~\cite{HoyosBadajoz:2009hb}, based on the description of ${\cal N}=2$ theories from D4 branes suspended between NS5 branes \cite{Witten:1997sc}. Flavor can be introduced by adding D6 branes. We will start with a configuration whose low energy limit on the T-dual D3 branes is ${\cal N}=4$ $U(2\Nc)$ super Yang-Mills plus $2 \Nf$ hypermultiplets in the fundamental representation, so the flavor group is $U(2\Nf)$.
In the T-dual configuration the flavor branes are D7's, and we will work in the 't Hooft limit of $\Nf/\Nc\ll 1$ so we can neglect their backreaction just as in the D3/D7 system of ref.~\cite{Karch:2002sh}. We then introduce an orientifold plane to produce an $SO(2 \Nc)$ theory with $USp(2 \Nf)$ flavor group and then finally do a $\mathbb{Z}_2$ orbifold projection that reduces it to $U(\Nc)$ with $U(\Nf)$ flavor group. We will show that an isospin chemical potential in the original $U(2 \Nc)$ theory is projected to a baryon chemical potential in the $U(\Nc)$ theory and discuss when the two theories are equivalent.

\subsection{Orientifold and orbifold projections}\label{sec:brane_configuration}
\hspace{0.51cm}

The construction in type IIA theory consists on a set of $2 \Nc$ D4 branes wrapping a circle in the $x^6$ direction and intersecting two $O6^+$ planes at opposite sides of the circle. In addition, there is a NS5 brane at each orientifold point and $2 \Nf$ D6 branes parallel to the $O6$ planes:
$$
\begin{array}{r|cccccccccc}
 \! & 0 & 1 & 2 & 3 & 4 &5 & 6& 7& 8 & 9 \\
\hline {\rm D4} & \times & \times & \times & \times& \cdot & \cdot & \times & \cdot & \cdot & \cdot \\
{\rm O6/D6} & \times & \times & \times & \times & \cdot & \cdot & \cdot & \times & \times & \times \\
{\rm NS5} & \times & \times & \times & \times & \times & \times & \cdot & \cdot & \cdot &\cdot
\end{array}
$$
Since the $O6$ planes are positively charged, Ramond-Ramond (RR) tadpoles do not cancel and the $\beta$ function for the 't Hooft coupling is positive. However, in the 't Hooft limit $\Nf\ll \Nc$, the $\beta$ function is suppressed by $\Nf/\Nc$ at large $\Nc$. So to leading order in $\Nf/\Nc$ we can neglect the tadpoles and consider the $D6$'s and $O6$'s as probes.

This brane setup has as a T-dual a configuration involving D3 and D7 branes. The two O6 planes map to a single O7 plane and the NS5 brane to a $\mathbb{Z}_2$ singularity localized at $x^6=x^7=x^8=x^9=0$:
$$
\begin{array}{r|cccccccccc}
 \! & 0 & 1 & 2 & 3 & 4 &5 & 6& 7& 8 & 9 \\
\hline {\rm D3} & \times & \times & \times & \times& \cdot & \cdot & \cdot & \cdot & \cdot & \cdot \\
{\rm O7/D7} & \times & \times & \times & \times & \cdot & \cdot & \times & \times & \times & \times \\
\mathbb{Z}_2 & \times & \times & \times & \times & \times & \times & \cdot & \cdot & \cdot & \cdot
\end{array}
$$
The geometric effect of the $\mathbb{Z}_2$ action is a reflection in the transverse directions. The orientifold projection $\Omega'=\Omega R_{45} (-1)^{F_L}$ involves worldsheet parity reversal $\Omega$, a reflection $R_{45}$ in the $x^4$ and $x^5$ coordinates, and $(-1)^{F_L}$ acts as $-1$ in the Ramond sector of left movers. The effect on Chan-Paton factors of open strings on D3 branes is given by the matrices
$\gamma_3 = i J_{2 \Nc}$ for the orbifold action and $\omega_3 = \textbf{1}_{2 \Nc}$ for the orientifold action. The corresponding matrices for the D7 branes are $\gamma_7 = i J_{2 \Nf}$ and $\omega_7 = i J_{2 \Nf}$.

The massless spectrum of D3 branes involves a vector multiplet on the worldvolume $A_{0123}$ and three complex scalar multiplets describing the transverse motion $X_{45}$, $X_{67}$, $X_{89}$. Before the projection those describe the field content of ${\cal N}=4$ $U(2 \Nc)$ super Yang-Mills, that in ${\cal N}=2$ language involves a vector multiplet and a hypermultiplet in the adjoint representation. The orientifold action is
\begin{equation}
\begin{array}{rcl}
A_{0123} & \to & -\omega_3 \; A_{0123}^T\; \omega_3^{-1}, \\[3pt]
X_{45} & \to & -\omega_3 \; X_{45}^T\; \omega_3^{-1} , \\[3pt]
X_{67,89} & \to & \omega_3 \; X_{67,89}^T\; \omega_3^{-1}.
\end{array}
\end{equation}
Therefore, the orientifold projection for the gauge field is
\begin{equation}\label{eq:projomega}
\cP_\omega A_\mu = \tfrac{1}{2}\left(A_\mu-A_\mu^T \right),
\end{equation}
so the projected gauge field is antisymmetric and spans an $SO(2 \Nc)$ algebra. The field $X_{45}$ is in an antisymmetric (adjoint) representation, while for the fields $X_{67,89}$ the orientifold action projects them to a symmetric representation.

The $\mathbb{Z}_2$ action of the orbifold is
\begin{equation}
\begin{array}{rcl}
\cP_\omega A_{0123} & \to &  \gamma_3\; \cP_\omega A_{0123}\; \gamma_3^{-1}, \\[3pt]
\cP_\omega X_{45} & \to &  \gamma_3\; \cP_\omega X_{45}\; \gamma_3^{-1}, \\[3pt]
\cP_\omega X_{67,89} & \to & -\gamma_3\; \cP_\omega X_{67,89}\; \gamma_3^{-1}.
\end{array}
\end{equation}
The transformations of $A_{0123}$ and $X_{45}$ are identical and produce fields in the adjoint representation of $U(\Nc)$. The projection on $X_{67,89}$ produces fields in a two-index symmetric representation. More explicitly, for the gauge field the projection is
\begin{equation}\label{eq:projgamma}
\cP_\gamma \cP_\omega A_\mu = \tfrac{1}{2}\left( \cP_\omega A_\mu+J_{2 \Nc}\cP_\omega A_\mu J_{2 \Nc}^{-1}\right).
\end{equation}
The resulting theory is a ${\cal N}=2$ $U(\Nc)$ theory with a symmetric hypermultiplet. If one considers the orientifold projection alone, the theory is projected to ${\cal N}= 2$ $SO(2 \Nc)$ super Yang-Mills with a hypermultiplet in the two-index representation, we can think of this theory as the analog of the $SO(2 \Nc)$ gauge theory of the QCD case.

The D3/D7 spectrum is initially described by two $2\Nc\times 2\Nf$ chiral multiplets $H^A$ describing strings from D3 to D7 branes and the reversed strings ${\widetilde H}_A=\epsilon_{AB} {H^B}^\dagger$. The orientifold and orbifold actions are as follows
\begin{equation}\label{eq:orientifoldproj}
\begin{array}{rclcrcl}
H^A & \to & - i \epsilon_{AB} \left(\omega_3 H^B \omega_7^{-1}\right)^*, & \  & \cP_\omega H^A &\to& \gamma_3 \cP_\omega H^A \gamma_7^{-1}.
\end{array}
\end{equation}
Therefore, the projections acting on flavor fields are
\begin{equation}\label{eq:projflav}
\cP_\omega H^A =\tfrac{1}{2}\left( H^A+ \epsilon_{AB} \left(H^B\right)^* J_{2 \Nf}^{-1}\right), \ \ \cP_\gamma\cP_\omega H^A = \tfrac{1}{2}\left(\cP_\omega H^A+J_{2 \Nc} \cP_\omega H^A J_{2 \Nf}^{-1}\right).
\end{equation}
The resulting massless field is a ${\cal N}=2$ hypermultiplet in the $(\Nc,\overline{\Nf})$ representation, or $\Nf$ flavors in the fundamental representation of the $U(\Nc)$ gauge group. In the theory obtained from the orientifold projection alone there are $\Nf$ hypermultiplets in the fundamental representation of the $SO(2 \Nc)$ gauge group. Although the maximal possible flavor group is $U(2 \Nf)$, in the theory at hand it is actually reduced to $USp(2\Nf)$, due to the coupling between the chiral components of the hypermultiplets with the chiral component of the vector multiplet in the superpotential
\begin{equation}
W \sim {\widetilde H} X H.
\end{equation}
Since $X$ is in the adjoint of $SO(2 \Nc)$, flavor indices in the superpotential are contracted with an antisymmetric form, which is invariant under a $USp(2 \Nf)\subset U(2 \Nf)$ subgroup.

The massless spectrum of D7 branes, that describes the BPS sector of flavored operators, is split between vector fields in the 0123 and 6789 directions, $A_{0123}$ and $A_{6789}$, and a scalar field in the 45 directions, $X_{45}$. Transformations act as
\begin{equation}\label{d7proj}
\begin{array}{rclcrcl}
A_{0123} & \to &  - \omega_7\; A_{0123}^T\; \omega_7^{-1}, &  &  \cP_\omega A_{0123}& \to & \gamma_7 \; \cP_\omega A_{0123}\; \gamma_7^{-1},\\[3pt]
X_{45} & \to & - \omega_7\; X_{45}^T\; \omega_7^{-1}, &  &  \cP_\omega X_{45} & \to & \gamma_7 \; \cP_\omega X_{45}\; \gamma_7^{-1}, \\[3pt]
A_{6789} & \to &  -\omega_7\; A_{6789}^T\; \omega_7^{-1}, &  &  \cP_\omega A_{6789} & \to & -\gamma_7 \; \cP_\omega A_{6789}\; \gamma_7^{-1}.
\end{array}
\end{equation}
Since the 8d Poincar\'e invariance is broken in the worldvolume of the D7 branes, the projection will be different for modes with dependence on the 6789 directions. The action \eqref{d7proj} for $A_{0123}$ and $X_{45}$ is valid for parity even modes while the action for $A_{6789}$ is valid for parity odd modes. This agrees with the $A_{0123}$ and $X_{45}$ components being scalar in the 6789 directions and $A_{6789}$ being a vector component. The flavor group is $U(\Nf)$, but the spectrum of BPS operators is different from the original theory since the hypermultiplet in the D3 sector is in the two-index symmetric representation and not in the adjoint.

The holographic dual description is type IIB string theory on $AdS_5\times \bR\bP^5$, with D7 probe branes that sit on top of O7 planes wrapping a $\bR\bP^3\subset \bR\bP^5$ cycle. The $AdS_5\times \bR\bP^5$ geometry can be understood using a different basis of transformations. The O7 action is $\Omega_7 = \Omega R_{45} (-1)^{F_L}$, while the $\mathbb{Z}_2$ singularity acts as a $R_{6789}$ reflection on the geometry. Since O7 planes and O3 planes have the same effect on Ramond forms (cf.~\cite{Hanany:2000fq}), the combined action is equivalent to the action of an O3 plane $\Omega_3=R_{6789}\Omega_7 = \Omega R_{456789} (-1)^{F_L}$. The action of the O3 plane on $AdS_5\times S^5$ is known to give the $\bR\bP^5$ geometry, since it acts as a reflection on the space transverse to the D3 branes \cite{Witten:1998xy}. From the T-dual perspective this geometry without the O7 orientifold can be constructed from a stack of D4 branes sitting on O4$^-$ or O4$^+$, giving holographic duals 
 with orthogonal or symplectic gauge groups.

\subsection{From isospin to baryon chemical potential}\label{sec:holographic_proof}
\hspace{0.51cm}

The dynamics of the probe D7 branes in the D3 background are determined by the DBI action,
\begin{eqnarray}\label{eq:DBI}
S_{DBI}=-T_7\int d^8\xi\,\ {\rm Tr} \sqrt{-\det\left(G+2\pi\alpha'F\right)},
\end{eqnarray}
where $\xi$ are the world-volume coordinates, $G$ is the pull-back of the spacetime metric to the world volume and $F$ is the field strength of the gauge fields on the brane. The $2 \Nf$ D7 branes in the $U(2 \Nc)$ theory are wrapping an $S^3\subset S^5$. Writing the $AdS_5\times S^5$ metric as
\begin{equation}
ds^2=\frac{|y|^2}{R^2}\, \eta_{\mu\nu} \, dx^\mu dx^\nu+\frac{R^2}{|y|^2}\sum_{i=4}^9 dy_i^2,
\end{equation}
the D7's are localized at $y_8=y_9=0$ and extend along all the other directions. As we have explained, the full projection identifies points in the geometry that map to each other under a reflection $y^i\to -y^i$.

An isospin chemical potential in the $U(2 \Nc)$ field theory is described by a background gauge field on the D7 brane
\begin{equation}\label{eq:isosp}
A_0=i \mu \,J_{2 \Nf}.
\end{equation}
More generally, this will be taken as the boundary condition for $A_0$. Notice that this configuration survives both projections \eqref{d7proj}. To highlight the effect of the projection let us write the gauge potential on the D7 brane as
\begin{eqnarray}
A_0=\left(
\begin{array}{cc}
H & C\\
C^\dagger & H'
\end{array}
\right),
\end{eqnarray}
where $H,H'$ and $C$ are $\Nf\times \Nf$ Hermitian and general complex matrices, respectively. The orientifold projection in \eqref{d7proj} implies the conditions
\begin{equation}\label{eq:cond1}
H'=-H^T, \ \ C=C^T,
\end{equation}
so $A_0$ is in the adjoint of a $USp(2 \Nf)$ group,
\begin{equation}
J_{2 \Nf} A_0 +A_0^T J_{2 \Nf}=0.
\end{equation}
The orbifold projection based on the transformation \eqref{d7proj} imposes
the conditions
\begin{equation}\label{eq:cond2}
H=-H^T, \ \ C=-C^*.
\end{equation}
So $H$ is reduced to a purely imaginary antisymmetric matrix and $C$ to a purely imaginary symmetric matrix. The combination $\tilde{A}_0=-iC-H$ belongs to the adjoint representation of a $U(\Nf)$ group. One can check this by doing a global transformation $A_0\to U A_0 U^\dagger$ with elements of the unbroken gauge group
\begin{equation}
U \omega_7 \, U^T=\omega_7,\ \  U\gamma_7 \, U^\dagger= \gamma_7.
\end{equation}
In terms of the $U(\Nf)$ gauge field, the configuration \eqref{eq:isosp} maps to
\begin{equation}\label{eq:baryop}
\tilde{A}_0=\mu\textbf{1}_{\Nf},
\end{equation}
which corresponds to a baryon chemical potential. This shows that indeed the isospin chemical potential is projected to a baryon chemical potential. Notice that after the orientifold projection, because the gauge group $SO(2\Nc)$ is real, the isospin  chemical potential is equivalent to the ``baryon number'' chemical potential.

\subsection{Validity of the equivalence}
\hspace{0.51cm}

The definition of the non-Abelian DBI action without derivatives is ambiguous,
as it is possible to use
$[D_\mu,D_\nu] F_{\alpha,\beta}=[F_{\mu\nu},F_{\alpha\beta}]$ to convert between
derivatives and field strengths.
Nevertheless,
a possible approach is to define the non-Abelian DBI action by giving an ordering prescription for the trace when the gradients of the field strength are small \cite{Tseytlin:1997csa}.
In this regime, the DBI action does not change under the projections.
We show this more explicitly in Appendix~\ref{sec:order}.
Under this assumption, we obtain the same results in the original and in the daughter theories, as long as all the components that are not invariant under the projection are zero.
Notice that if restricted to questions about the ground state configuration with or without charge density, there are no ordering ambiguities; the configuration is Abelian so the separation between derivatives and field strengths is well-defined and any ordering prescription leads to the same results.

From the perspective of the dual field theory, this means that the parent and the daughter theories are equivalent in the large-$\Nc$ limit.
In other words, the equivalence between the original and the daughter theory holds as long as the solution to the equations of motion obtained from the DBI action and the boundary conditions are invariant under the projection symmetry.
We can fix the boundary conditions, but there is still the possibility that the correct solution breaks spontaneously one of the symmetries we have used to define the projection.
This is a dynamical question, but fortunately one that can be answered in this context.

\begin{figure}[t]
\begin{center}
\includegraphics[scale=0.4]{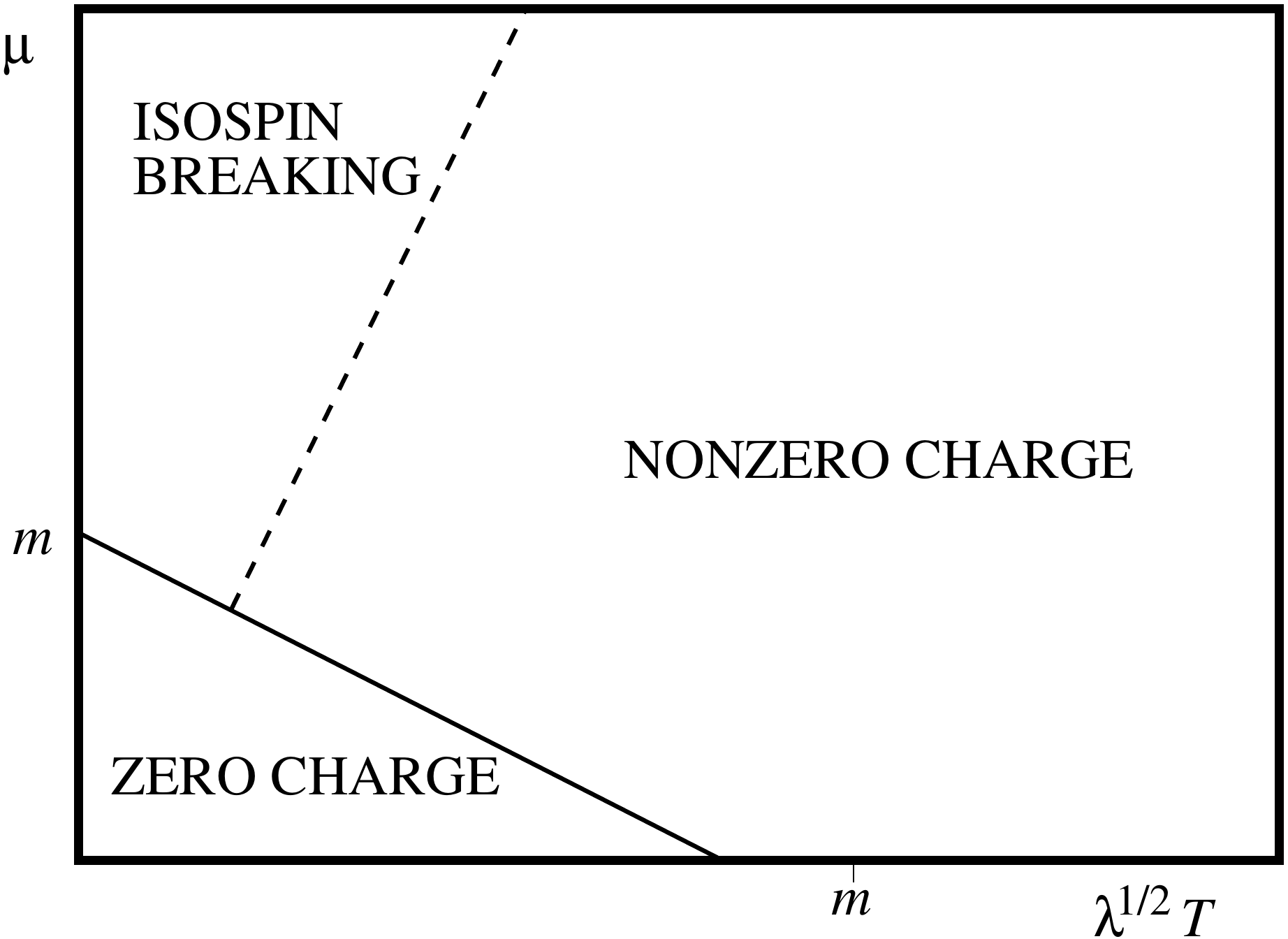}
\caption{
Sketch of the phase diagram in the chemical potential $\mu$ and temperature $T$ plane,
for either baryon or isospin chemical potential.
The phase with spontaneous isospin breaking only exists for
the case of an isospin chemical potential.
The large $\Nc$ equivalence relating baryon and isospin chemical potential is valid
in the confined and deconfined phases, but not in the region where isospin can be
spontaneously broken.
}\label{fig:PhaseDiagram}
\vspace*{-12pt}
\end{center}
\end{figure}

The phase diagram of D7 branes in the D3 background geometry was studied in ref.~\cite{Mateos:2007vc} for a baryon chemical potential,
and also in refs. \cite{Erdmenger:2008yj,Ammon:2008fc,Erdmenger:2011hp}
for an isospin chemical potential.%
\footnote{%
Notice that we are considering a baryon chemical potential in a theory where two of the adjoint fields have been changed to two-index symmetric representations. However, both theories are equivalent in the large-$\Nc$ limit. In the gravity side we have geometries with the same metric but different topologies, so solutions to the classical equations of motion are the same in both cases.
}
We have sketched the phase diagram in Figure~\ref{fig:PhaseDiagram}. For non-zero quark mass $m$ there is a phase with no charge density, where quarks can form bound states of mass $\sim m/\sqrt{\lambda}$ and there is a discrete spectrum of low-spin mesonic states. For large enough chemical potential and/or temperature there is a phase transition,%
\footnote{%
The phase transition is first order at zero chemical potential, but changes to 3rd order at a tricritical point  \cite{Faulkner:2008hm} at finite T and $\mu$. At zero temperature the transition becomes second order \cite{Karch:2007br}.}
and a finite charge density (baryon or isospin) appears.
In this phase the quarks do not form bound states and the spectrum is continuous.
The curves of the phase transition in the $(\mu_{B,I},T)$ plane are the same for both baryon and isospin chemical potential, as we expect from the equivalence.
The equivalence also holds in the finite density phase,\footnote{%
In ref.~\cite{Erdmenger:2008yj} it was observed that there is a 4-fold symmetry in the $(\mu_B,\mu_I)$ plane.
}
but for an isospin chemical potential there is also a symmetry broken phase where vector mesons charged under the $U(1)$ isospin symmetry condense \cite{Ammon:2008fc,Basu:2008bh}. The broken phase appears for values of the chemical potential larger than the scales given by the temperature and the mass of the mesons, which is proportional to the quark mass in these models. This implies in particular that at zero temperature and in the chiral limit the equivalence will fail. However, for massive quarks and small chemical potentials the equivalence will still be valid.

\subsection{$\Nf/\Nc$ corrections}
\hspace{0.51cm}

It is interesting to study the question of $\Nf/\Nc$ corrections in the theories with holographic duals and compare with QCD-like theories.
Up to now, we have been working in a probe approximation for the flavor branes. From the field theory perspective this is equivalent to working in the 't Hooft limit of $\Nf/\Nc\to 0$, with no quark loops in planar diagrams.
To find the effect of $\Nf/\Nc$ corrections, one should compute the backreaction of the brane on the geometry.
In principle, orientifold planes also produce a backreaction on the geometry of the same order; from the point of view of the field theory dual this corresponds to $1/\Nc$ corrections not associated with flavor.
However, in the limit where the number of flavor branes is large, but still much smaller than the number of color branes, $\Nc\gg \Nf \gg 1$, one can neglect the backreaction of the orientifold to leading order.
From the field theory perspective, one can do a double expansion in $\Nf/\Nc$ and
$1/\Nc$, where $\Nf/\Nc$ corrections appear in planar diagrams with any number of flavor loops and $1/\Nc$ corrections correspond to non-planar diagrams.

The comparison of $\Nf/\Nc$ corrections is a bit subtle. On the one hand, the DBI action is the same for both the normal configuration dual to isospin chemical potential,
and the orientifolded configuration dual to baryon chemical potential.
Wess-Zumino terms with an even number of field strengths are also the identical.
However, terms with an odd number of field strengths can differ.
Consider, for instance, the coupling of the D7 to the RR potential $C_6$,
\begin{equation}
S_{WZ} =\mu_7 \int_{D7} C_6\wedge \tr(F).
\end{equation}
Before introducing the orientifold, $\tr(F)=0$, since the gauge field on the brane is non-Abelian \eqref{eq:isosp}. The projection does not affect to the value of the trace, but for a D7 brane with an Abelian configuration \eqref{eq:baryop}, $\tr(F)\neq 0$ in general, so this term is different in both cases and the backreaction of the D7 branes will also be different. Therefore, $\Nf/\Nc$ corrections in the parent and daughter theories will be different in general and the equivalence only holds in the strict 't Hooft limit.

From the field theory perspective, we saw in section \eqref{sec:qcdcomments} that the condition \eqref{eq:regularity} is not satisfied in the projection to a baryon chemical potential. For the supersymmetric theory we have to check two projections: both the orbifold projection, and the orientifold projection.
The orbifold projection $\cP_\gamma$ mapping $SO(2 \Nc)\to U( \Nc)$,
whose action is given eqs.~\eqref{eq:projgamma} and \eqref{eq:projflav},
na\i\" vely seems to be regular.
However, the projection is written in a basis where fundamental fields are in a real representation, in particular it is valid for Majorana fermions, while in the daughter theory fermions will be Dirac.
In terms of a complex representation with Dirac fermions one recovers the same kind of projection as in \eqref{eq:qcdbarproj}, which is not regular.
Therefore, we also find from the field theory side that the equivalence will not hold for $\Nf/\Nc$ corrections.

One can show indirectly that the orientifold projection $\cP_\omega$ mapping
$U(2 \Nc)\to SO(2 \Nc)$ in eqs.~\eqref{eq:projomega} and \eqref{eq:projflav} is regular.
For this, notice that there is a regular projection from $SO(4 \Nc)\to U(2 \Nc)$ with isospin chemical potential using a $\mathbb{Z}_4$ subgroup.
This projection can be extended to $SO(4 \Nc)\to SO(2 \Nc)$ using a $D_4$ dihedral subgroup, with $\mathbb{Z}_4\subset D_4$. Indeed, using the orbifold action $A_\mu \to J_{4\Nc} A_\mu J_{4 \Nc}^{-1}$ to project $SO(4 \Nc)\to U(2 \Nc)$ and
\begin{equation}
L_{4 \Nc}=\left(\begin{array}{cc}
 \textbf{1}_{2 \Nc} & 0\\
0 & -\textbf{1}_{2 \Nc}
\end{array}
\right),
\end{equation}
to further project to $SO(2 \Nc)$ through the action $A_\mu \to L_{4 \Nc} A_\mu L_{4 \Nc}^{-1}$, one generates a regular representation of $D_4$.\footnote{The elements of the group are $\{\textbf{1}_{4 \Nc},-\textbf{1}_{4 \Nc},J_{4 \Nc},-J_{4 \Nc},L_{4 \Nc},-L_{4 \Nc},J_{4 \Nc} L_{4 \Nc}, L_{4 \Nc} J_{4 \Nc}\}$.} We can then group flavor fields in a $4\Nc\times 4 \Nf$ real matrix
\begin{equation}
\cH=\left(\begin{array}{cc}
 H_1+H_1^* & H_2+H_2^*\\
i(H_1^*-H_1) & i(H_2^*-H_2)
\end{array}
\right),
\end{equation}
and perform the orientifold projection \eqref{eq:projflav} in this basis as
\begin{equation}
\cP_\omega\cH=\tfrac{1}{2}\left(\cH+ L_{4 \Nc} \cH \Omega_{4 \Nf}^{-1}\right),
\end{equation}
where
\begin{equation}
\Omega_{4 \Nf}^{-1}=\left(\begin{array}{cc}
 0 & -J_{2 \Nf}^{-1}\\
J_{2 \Nf}^{-1} & 0
\end{array}
\right).
\end{equation}
Since $\tr \Omega_{4 \Nf}=0$ and $\Omega_{4 \Nf}^2=\textbf{1}_{4 \Nf}$, this shows that the orientifold projection in the flavor sector is also regular. Notice that the projection acting on fields on the flavor D7 branes is determined by $\Omega_{4 \Nf}$, so from the perspective of the field theory on the D7 branes this is a regular projection. In the holographic duals to both the parent $U(2 \Nc)$ and daughter $SO(2 \Nc)$ theories the gauge field configuration on the D7 brane is traceless, so the issue of terms with an odd number of field strengths does not arise in this case.

\section{Conclusion}\label{sec:conclusions}
\hspace{0.51cm}
We have used a holographic construction to demonstrate a large-$\Nc$ equivalence between theories with baryon and isospin chemical potential in the 't Hooft limit.
The equivalence is valid in the region of the phase diagram where neither isospin nor baryon symmetry are spontaneously broken.
The allowed region contains
a small temperature/chemical potential phase with no charge density and mesonic bound states,  and a large temperature/chemical potential phase with a finite charge density and a continuous spectrum.
Although the charge density vanishes in the low temperature phase in the classical supergravity approximation, it will be nonzero when one takes Hawking
radiation from the black hole into account.
From the perspective of the dual field theory, this indicates that any charge density is suppressed in the large-$\Nc$ limit, compared to the charge density that is present in the high temperature phase.
In other words, in the low temperature phase
the equivalence we have presented relates the leading order large-$\Nc$
behavior of ``vacuum'' thermodynamic properties of the theory with either
baryon or isospin chemical potential, but it does not provide information about
other details, such as properties of the thermal gas of mesons and baryons,
which are $1/\Nc$ suppressed.
The large $\Nc$ equivalence does relate non-trivial dependence on temperature and
chemical potential in the deconfined phase (with unbroken isospin and baryon number),
where there is temperature and chemical potential dependence at leading order in
$\Nc$.

Accepting the validity of gauge/string duality,
our analysis suggests that large $\Nc$ equivalences relating
theories with differing chemical potentials may be valid more generally
(when appropriate symmetry realizations hold).
Exactly the same projections which we have used for supersymmetric theories
can also be used to relate
$U(2 \Nc)$ QCD with $2 \Nf$ flavors and isospin chemical potential
to $U(\Nc)$ QCD with $\Nf$ flavors and baryon chemical potential.
However, because there is no known gravitational dual of QCD,
different methods are needed to construct a
purely field theoretic proof which applies to this case.
We hope to revisit these points in the future.

\section*{Acknowledgments}
We would like to thank Johanna Erdmenger and Patrick Kerner for stimulating discussions and comments.
This work was supported in part by the U.S. Department of Energy under Grant No. DE-FG02-96ER40956.
The work of M.H. is supported by Japan Society for the Promotion of Science Postdoctoral Fellowship for Research Abroad.

\appendix

\section{A quick introduction to the sign problem}\label{app:sign}
\hspace{0.51cm}
Consider pure Yang-Mills theory.
In lattice Monte-Carlo simulations, one generates field configurations with probability weight $e^{-S_{YM}}/Z_{YM}$ (times Haar measure),
where $S_{YM}$ is the Euclidean action and the partition function
$Z_{YM}\equiv \int dA_\mu e^{-S_{YM}[A]}$.
Expectation values
are approximated by taking the average over many configurations
generated by a Markov chain:
\begin{eqnarray}
\langle {\cal O}\rangle
\equiv
\lim_{k\to\infty}\frac{1}{k}\sum_{i=1}^k
{\cal O}[A_\mu^{(i)}]
=
\frac{1}{Z_{YM}}\int dA_\mu \> {\cal O}[A]\, e^{-S_{YM}[A]} \,.
\end{eqnarray}
Here $k$ is the number of lattice field configurations generated in the simulation
and $i=1,\cdots,k$ is a label distinguishing them.
In the sequence of configurations,
more likely configurations appear more often;
this is known as the ``importance sampling".

The key assumption, which is valid for pure gauge theories, is that the
the weight $e^{-S_{YM}}/Z_{YM}$ is \emph{real and positive},
so that the integrand of the functional integral 
may be regarded as a probability measure.
This condition can be broken when there are fermions in the theory.

To deal with fermions in a lattice theory, one performs the Grassmann integral
by hand.
For example,
the expectation value of the chiral condensate $\bar{\psi}\psi$ can be
expressed as
\begin{eqnarray}
\langle\bar{\psi}\psi\rangle
=
\frac{\int dA_\mu \> Tr\Slash{D}^{-1}[A]\cdot\det\Slash{D}[A]\cdot e^{-S_{YM}[A]}}{
\int dA_\mu \, \det\Slash{D}[A]\cdot e^{-S_{YM}[A]}} \,,
\end{eqnarray}
where $\Slash{D}$ denotes the lattice Dirac operator.
If $\det\Slash{D}[A]$ is real and positive for any gauge field $A_\mu$,
then one may simulate this system by using the effective action
$S_{eff}[A]=S_{YM}[A]-\log\det\Slash{D}[A]$.
However, the determinant $\det\Slash{D}[A]$ can, in general, be complex,
and then standard  Monte-Carlo techniques cannot be applied.
This is the so-called ``sign problem" (or  more properly ``phase problem"),
the word ``sign" referring to a possible negative sign of the determinant.
QCD with baryon chemical potential suffers from the sign problem, while
QCD with isospin chemical potential and the $SO(2\Nc)$ gauge theory with
fermion chemical potential are ``sign-free''.

One standard approach for dealing with the sign problem is the so-called
``reweighing'' method.
Consider the {\it phase quenched} ensemble
with weight $|\det\Slash{D}|e^{-S_{YM}}$. When the number of flavors is even,
the phase-quenched version of QCD with baryon chemical potential
is identical to QCD with an isospin chemical potential.
If $\langle \cdots \rangle_B$ denotes expectations in QCD with
a baryon chemical potential, and $\langle \cdots \rangle_I$ expectations
with an isospin potential, 
then it is immediate that
\begin{eqnarray}
\langle {\cal O} \rangle_B
=
\frac{\langle {\cal O}\cdot e^{i \eta} \rangle_I}{ \langle e^{i \eta} \rangle_I },
\label{eq:reweight}
\end{eqnarray}
where $e^{i\eta}$ is the phase of the fermion determinant
in the presence of a baryon chemical potential,
$e^{i\eta} \equiv \det\Slash{D}/|\det\Slash{D}|$.
Because one can apply standard Monte-Carlo simulation techniques to the
phase-quenched ensemble,
one can in principle evaluate the expectation value in the full theory
by computing both numerator and denominator in the identity (\ref{eq:reweight});
this is the reweighing method.
However, in practice reweighing works only when the phase does not
fluctuate violently.
Phase fluctuations grow
as the chemical potential $\mu_B$ is increased,
and as the lattice volume $V$ grows.
The logarithm of the fermion determinant is extensive,
$\ln \det \Slash D = O(V)$,
and (with a non-zero baryon chemical potential),
so is its imaginary part, $\eta$.
This implies that both numerator and denominator of eq.~(\ref{eq:reweight})
vanish exponentially in the thermodynamic limit, making their estimation
via Monte Carlo methods increasingly problematic as the volume $V \to \infty$.

\section{Projection of the DBI action}\label{sec:order}
\hspace{0.51cm}
The DBI action of a $D(d{-}1)$ brane may be expressed as
\begin{eqnarray}\label{eq:DBI2}
S_{DBI}=-T_{d-1}\int d^d\xi\,\ \, {\rm \mathbf{S'Tr}}_{2 \Nf} \sqrt{-\det\left(G+2\pi\alpha'F\right)},
\end{eqnarray}
where we use the notation ${\rm \mathbf{S'Tr}}_{2 \Nf}$ for a trace of $2 \Nf\times 2 \Nf$ matrices with a predetermined ordering prescription that is, however, unknown.
Up to $F^4$ terms,
it should coincide with the symmetrized trace prescription \cite{Tseytlin:1997csa}.

For a $d$-dimensional brane, one can write the determinant as
\begin{equation}\label{eq:expansion0}
 \det\left(G+2\pi\alpha'F\right)=\tfrac{1}{d!}\, \epsilon^{\mu_1\cdots\mu_d}\epsilon^{\nu_1\cdots\nu_d} (G_{\mu_1\nu_1}+2\pi \alpha' F_{\mu_1\nu_1})\cdots  (G_{\mu_d\nu_d}+2\pi \alpha' F_{\mu_d\nu_d})\,,
\end{equation}
where $G_{\mu_1\nu_1}$ is the pullback of the metric and is proportional to the identity matrix. The field strengths $F_{\mu_1\nu_1}$ are proportional to the generators of the gauge group on the brane. Since this is a matrix product one has to define the order, we will not assume a particular ordering in the following. We can extract the metric factors as
\begin{align}\label{eq:expansion1}
 &\det\left(G+2\pi\alpha'F\right)
\nonumber\\&\qquad{}
 =\tfrac{1}{d!}\, \epsilon^{\mu_1\cdots\mu_d}\epsilon^{\nu_1\cdots\nu_d}G_{\mu_1\alpha_1}\cdots G_{\mu_d\alpha_d}(\delta^{\alpha_1}_{\nu_1} +2\pi \alpha' G^{\alpha_1\beta_1} F_{\beta_1\nu_1})\cdots  (\delta^{\alpha_d}_{\nu_d} +2\pi \alpha' G^{\alpha_d\beta_d} F_{\beta_d\nu_d})
\nonumber\\&\qquad{}
 =\tfrac{1}{d!}\,(\det G)\, \epsilon^{\mu_1\cdots\mu_d}\epsilon_{\alpha_1\cdots\alpha_d}(\delta^{\alpha_1}_{\nu_1} +2\pi \alpha' G^{\alpha_1\beta_1} F_{\beta_1\nu_1})\cdots  (\delta^{\alpha_d}_{\nu_d} +2\pi \alpha' G^{\alpha_d\beta_d} F_{\beta_d\nu_d})\,.
\end{align}
Alternatively, one can write the determinant as
\begin{align}\label{eq:expansion2}
 &\det\left(G+2\pi\alpha'F\right)
\nonumber\\&\qquad{}
 =\tfrac{1}{d!}\, \epsilon^{\mu_1\cdots\mu_d}\epsilon^{\nu_1\cdots\nu_d}G_{\alpha_1\nu_1}\cdots G_{\alpha_d\nu_d}(\delta^{\alpha_1}_{\mu_1} +2\pi \alpha' G^{\alpha_1\beta_1} F_{\mu_1\beta_1})\cdots  (\delta^{\alpha_d}_{\mu_d} +2\pi \alpha' G^{\alpha_d\beta_d} F_{\mu_d\beta_d})
\nonumber\\&\qquad{}
 =\tfrac{1}{d!}\, (\det G)\, \epsilon^{\nu_1\cdots\nu_d}\epsilon_{\alpha_1\cdots\alpha_d}(\delta^{\alpha_1}_{\mu_1} +2\pi \alpha' G^{\alpha_1\beta_1} F_{\mu_1\beta_1})\cdots  (\delta^{\alpha_d}_{\mu_d} +2\pi \alpha' G^{\alpha_d\beta_d} F_{\mu_d\beta_d})\,.
\end{align}
Since
$
G^{\alpha\beta} F_{\beta\nu}=-G^{\alpha\beta} F_{\nu\beta}
$,
after relabeling indices $\nu_i\leftrightarrow\mu_i$ in \eqref{eq:expansion2} and comparing with \eqref{eq:expansion1}, one finds that
$
 \det\left(G+2\pi\alpha'F\right)= \det\left(G-2\pi\alpha'F\right)
$.
Therefore, only even powers of the field strength $F$ appear in the expansion of the determinant, which can be seen as a manifestation of charge conjugation invariance.
Taking the square root and defining $\sigma_N(T^N)$ as a possible ordering of $N$ generators $T$ appearing in the trace,\footnote{Since $\sigma_N$ is defined for a trace, it maps to an element of the quotient of the permutation group over the cyclic group $S_n/C_n$, which corresponds to the conjugacy classes of $S_n$.
These can be classified using Young tableaux.} we have
\begin{align}\label{eq:expansion}
&{\rm \mathbf{S'Tr}}_{2 \Nf}\sqrt{-\det\left(G+2\pi\alpha'F\right)}
= \sqrt{-\det G}
\nonumber\\&{}\times
\left[2 \Nf+ \sum_{N\geq 1} (\alpha')^{2 N}\sum_{\sigma_{2N}} \sum_{k\geq 1} \left(\prod_{q=1}^k \sum_{n_q=0}^{[d/2]} \right) \delta_{\sum_q n_q-N}\, c^{N,k}_{\ \ n_1,n_2,\cdots,n_k} \tr \sigma_{2N}\left(F^{2 n_1} F^{2 n_2} \cdots F^{2 n_k}\right) \right].
\end{align}
Where we have suppressed spacetime indices and denote $F^n$ as a product of $n$ field strengths appearing in the determinant. The largest possible power is $d$ if $d$ is even or $d-1$ if $d$ is odd. The action has to be Hermitian, this implies that given some ordering $\sigma_{2N}$, the reversed ordering $\sigma_{2N}^T$ also appears with the same coefficients. For instance, if we have $\tr\left(F_1\, F_2\,\cdots\, F_{n-1}\, F_n\right)$, the Hermitian conjugate is $\tr\left(F_n\, F_{n-1}\,\cdots\, F_2\, F_1\right)$.

Although expression \eqref{eq:expansion}  takes the form of an $\alpha'$ expansion, we are overlooking other $\alpha'$ corrections involving derivatives of the field strength and  $\alpha'$ corrections that depend on the background metric.
The former can formally be included by allowing $DF$ factors inside the traces; many new terms (with unknown coefficients) would appear but the basic structure would not change if discrete symmetries (C,P,T) are not broken.
We will just assume that gradients of the field strength are much smaller than their magnitude.
Corrections to the geometry are more problematic, as $\alpha'$ corrections are,
in general, different in the presence of an orientifold plane.
But since we are both in the supergravity approximation and in a probe limit,
we can neglect these corrections so the coefficients of the expansion are not
affected.
Under these assumptions the DBI action of the daughter theory is the
na\"{\i}ve projection of the original theory, as we will now show.

Let us write the field strength of the gauge field on the D-brane as
\begin{eqnarray}
F=\left(
\begin{array}{cc}
H & C\\
C^\dagger & H'
\end{array}\right).
\end{eqnarray}
Here and in the following, we suppress spacetime indices.
Under a global $U(2 \Nf)$ transformation taking $F\to P FP^{-1}$,
with $P$ defined in \eqref{eq:P}, the field strength transforms to
\begin{eqnarray}
PFP^{-1}=\frac{1}{2}\left(
\begin{array}{cc}
H+H'+i(C^\dagger-C) & H-H'+i(C^\dagger+C)\\
H-H'-i(C+C^\dagger) & H+H'-i(C^\dagger-C)
\end{array}\right).
\end{eqnarray}
After applying the conditions \eqref{eq:cond1} and \eqref{eq:cond2} which
follow from the projection,
the transformed field strength is block-diagonal,
\begin{eqnarray}
PFP^{-1}=\left(
\begin{array}{cc}
H-iC & 0\\
0 & H+i C
\end{array}\right)=\left(
\begin{array}{cc}
{\tilde F}^T & 0\\
0 & -{\tilde F}
\end{array}\right).
\end{eqnarray}
Here $\tilde{F}$ is the $U(\Nf)$ gauge field of the daughter theory. Any power of the field strength $F^n$ has the same block-diagonal form after the global transformation, so for $\sum_k \ell_k=L$ factors, the trace is
\begin{align}
\tr\sigma_L\left(\prod_k F^{\ell_k}\right)
&{}=\tr\sigma_L\left(P \prod_k F^{\ell_k} P^{-1}\right)
=\tr\sigma_L\left(\prod_k (\tilde{F}^T)^{\ell_k}\right)+\tr\sigma_L\left(\prod_k (-\tilde{F})^{\ell_k}\right)
\nonumber\\&{}
= \tr\sigma_L^T\left(\prod_k \tilde{F}^{\ell_k}\right)+(-1)^L \tr\sigma_L \left(\prod_k \tilde{F}^{\ell_k}\right).
\end{align}
Since $L$ is always even in the expansion \eqref{eq:expansion}, the phase factor is trivial $(-1)^L=(-1)^{2N}=1$. For each ordering $\sigma_{2N}$ in \eqref{eq:expansion} we get two terms, one corresponding to the same ordering for the $U(\Nf)$ gauge field and another one corresponding to the reversed order. From the Hermiticity of the action in the $U(2 \Nf)$ theory, we should have another contribution in the projected action that is exactly the same but whose origin is a term with reversed order $\sigma_{2N}^T$. Adding the two together,
we have a Hermitian action where the traces are projected to
\begin{equation}
\tr \sigma_{2N}\left(F^{2 n_1} F^{2 n_2} \cdots F^{2 n_k}\right)\to 2\tr \sigma_{2N}\left(\tilde{F}^{2 n_1} \tilde{F}^{2 n_2} \cdots \tilde{F}^{2 n_k}\right).
\end{equation}
Finally, using expression \eqref{eq:expansion},
\begin{equation}
{\rm \mathbf{S'Tr}}_{2 \Nf}\sqrt{-\det\left(G+2\pi\alpha'F\right)}
\to
2\, {\rm \mathbf{S'Tr}}_{\Nf}\sqrt{-\det\left(G+2\pi\alpha'\tilde{F}\right)}
\end{equation}
This proves the equivalence of the ordering prescription in the original and the projected actions. Comparing with the action \eqref{eq:DBI2}, the tension of the D-brane  in the $U(2 \Nf)$ theory is half the tension of the $U(\Nf)$ theory. However, this is compensated by the volume of the internal space, that is halved when we project from the $S^5$ to the ${\bf RP}^5$ geometry.


\end{document}